\documentstyle[aps,epsfig,twocolumn]{revtex}

\begin{document}

\draft  

\title{Dynamic splitting of a Bose-Einstein condensate}

\author{ C.~Menotti $^{1,2,3}$, J.R.~Anglin$^{4}$, J.I.~Cirac$^1$ and P.~Zoller$^1$}
\address{ $1$  Institut f\"ur Theoretische Physik, Universit\"at Innsbruck,
                A--6020 Innsbruck, Austria  }

\address{ $2$  Scuola Normale Superiore, I--56126 Pisa, Italy }

\address{ $3$   Unit\`a INFM, Dipartimento di Fisica, Universit\`a di Pisa, I--56126 
Pisa, Italy }

\address{ $4$ Institute for Theoretical Atomic and Molecular Physics, \\
Harvard-Smithsonian Center for Astrophysics, Cambridge MA 02135}

\pagestyle{headings}
\maketitle

\begin{abstract}
We study the dynamic process of splitting a condensate by raising a potential 
barrier in the center  of a harmonic trap. We use a two-mode model to describe 
the phase  coherence between the two halves of the condensate. Furthermore,
we explicitly consider the spatial dependence of the mode funtions, which
varies depending on the potential barrier. This allows to get the
tunneling coupling between the two wells and  the on-site energy as a function 
of the barrier height. Moreover we can get some insight on the collective modes 
which are excited by  raising  the barrier.
We describe the internal and external degrees of freedom by variational ansatz.
We  distinguish the possible regimes as a function of the characteristic 
parameters of the problem and  identify the adiabaticity conditions.
\end{abstract}


\section{Introduction}

Bose-Einstein condensation of an ideal gas is typically presented in 
introductory textbooks solely in terms of particle numbers.  And quantum 
mechanically enhanced number densities were the `smoking gun' observed in the 
first experiments on dilute gas condensates.  But many phenomena of interacting 
condensates depend critically on the conjugate quantity to particle number, 
namely the quantum mechanical phase 
\cite{prl77-2159,prl77-3489,jourmodopt44-1775,EPJD8-319,pra58-531}.  
One can have highly occupied states with or without phase coherence 
between them, and the presence or absence of phase coherence can make a dramatic 
difference in the physical properties of an ultra-cold gas.  As in the case of a 
gas held in an optical lattice, which can be a superfluid or a Mott insulator 
depending on the strength of the lattice, the onset or loss of phase coherence 
can even be a phase transition \cite{prl81-3108}.

Since the global phase of a condensate is unobservable, the 
simplest system in which phase coherence can be manifested consists of two 
states, occupied by a large number of bosons.  As such a system can 
realistically be approximated by a condensate in a double well, it has recently 
attracted attention 
\cite{pra55-4318,prl79-4950,pra57-r28,condmat9905059,prl78-4675,prl81-1344,prl81-1345,pra58-566,condmat9812260}.
In these works, Josephson oscillations and the phase coherence between 
two coupled condensate were studied considering time independent coupling parameters. 
In \cite{pra60-2351} the disappearance of the phase coherence between the two wells due
to a change in time of the tunneling coupling has been studied.
The coupling parameters and the related phase coherence 
properties depend on the potential barrier between the two wells,
since in the limit where the barrier  is very low we have a single
condensate and in the limit where the barrier is very high we have two completely
separated condensates. Even if, at least in the high barrier limit, it has often been 
pointed out how to relate the coupling parameters (on-site energy and tunneling 
coupling) to the overlap of  the wavefunctions localised in the two wells, 
this has never really been taken explicitly into account.
Introducing the spatial degrees of freedom, as we did, allows us to relate all that to 
the potential  barrier in a more than phenomenological way and
becomes expecially important in the study 
of the dynamics of the process, because the wavefunctions change drastically in time
and collective modes are excited.

In this paper we examine this problem and develop a method which can be 
extended to the case of many wells, in order to encompass the turning on of 
an optical lattice in a condensate, allowing to go beyond  a Gross-Pitaevkii 
treatment, as the one done for example in \cite{pra58-1480}.
The physical situation is similar to the ones already realized experimentally,
where a double well potential was created by shining a far-off resonant
laser beam in the center of the magnetic trap (see e.g. \cite{ketterle})
or where an array of traps was created by on optical standing wave 
\cite{science82-1686}. 

\begin{figure}[htb]
\begin{center}
\epsfig{file=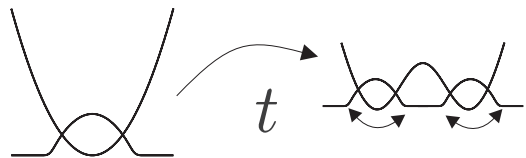,width=0.9\linewidth}
\caption{Schematic diagram of the process.}
\end{center}
\end{figure}

We are interested in the full dynamics of the process: in addition to the
phase coherence properties, we want to study  the excitation of the collective
modes by taking the spatial dependence of the condensate
wavefunction explicitly into account.  The price of including all these effects 
without assuming mean field theory is that we must use a time-dependent 
variational approach, choosing variational
ansatz to describe both the ``internal'' and ``external'' dynamics 
(that is, the distribution of particles between two motional 
states treated as given, and the evolution of the spatial wave functions of 
these states).  
This allows us to reduce the intractable full problem to a set of coupled 
differential equation for our few variational parameters. 
Although the time-dependent 
variational approach is not guaranteed to be quantitatively accurate, it allows 
qualitatively important processes to be investigated, and it has proven 
surprisingly reliable in previous applications to condensate physics
\cite{prl77-5320}.  
Here we  use it to derive the coupling between
the internal and external dynamics, investigate which are the conditions
under which the two can be decoupled, and identify the typical time scales
for both.

In Sec.~\ref{Model} we briefly present our two-state model and discuss 
qualitatively the
behaviour of the system that our model can explain. In Sec.~\ref{ansatz},
we introduce explicitly the variational ansatz we choose to describe
the phase dynamics and the collective modes. After introducing the 
time--dependent variational principle and  deriving the Lagrangian,
we write the equations of motion for the variational parameters. Then 
in Sec.~\ref{decouple} we discuss the conditions under which it is an accurate 
approximation to neglect the coupling between internal and external degrees 
of freedom. 
This decoupling allows us to model both internal and external evolution in a 
still simpler way. The external
dynamics is studied in Sec.~\ref{finexc}, comparing simple analytic
models with numerical results. Sec.~\ref{numres} is devoted to the internal 
dynamics, where the statics is studied and analytic estimates obtained.
With these decoupled studies to guide expectations, the full variational 
equations of motion, with coupled internal and external degrees of freedom, will 
be studied numerically in
Sec.~\ref{results}.  The results are compared with those of the phase model 
describing the relative phase of two weakly coupled superfluids
\cite{foundphys21-353,physb194-196-1389}.
  We conclude with a general discussion 
of our results and their implications.   Two appendices define the number 
difference and relative phase operators
, derive the phase model Hamiltonian,
and show that our variational ansatz adequately represents the evolution 
generated by it.

\section{Model}\label{Model}

Let us consider the situation in which we have $N$ bosons
confined in a harmonic trap at zero temperature. We slowly
deform the trap symmetrically around its center raising a potential
barrier, until it becomes
a double--well potential. We want to study the dynamics of the
process as well as the final state of the bosons.

We will treat the problem in one dimension. This
describes the situation in which we have a cigar-shaped
condensate, and we deform it along the most elongated direction.
Nevertheless, our model could be extendend in a straightforward
way to $2$ and $3$ dimensions. This would just lead to more
complicated equations, but would not affect the main results.

\subsection{Two--mode model}

The second quantization Hamiltonian describing the situation we
have in mind and in which particles interact via a
$\delta$-pseudopotential is

\begin{eqnarray} \label{1dh}
\hat{H}(t)&=&\int \hat{\Psi}^{\dag}(z)
\left( \frac{p^2}{2M} + V(z,t) \right) \hat{\Psi}(z) dz+ \nonumber      \\
&+&
\frac{1}{2}\;g
\int \hat{\Psi}^{\dag}(z)  \hat{\Psi}^{\dag}(z)
  \hat{\Psi}(z)  \hat{\Psi}(z) dz,
\end{eqnarray}
where $\hat{\Psi}$ is the bosonic field operator and
$V(z,t)$ is the time--dependent potential which describes the
deformed trap. Here, $g$ is an effective coupling constant which
depends both on the $s$-wave scattering length and on the atomic
distribution in the transverse directions.

We assume that at time $t=0$ the atoms are in the ground state of
this Hamiltonian, and we want to determine the state after the
potential is deformed. This problem cannot be solved even
numerically (see, for example, \cite{condmat0003399} and references
therein). Therefore, we need to consider a simplified model
which describes the main features of the process.  
If the process is completely adiabatic, the final state
will be a `fragmented condensate' with half of the particles in each of the 
potential wells: when the two condensates do not interact 
this is a much lower energy state than the one with phase coherence, 
because minimizing fluctuations in the relative particle number
lowers the energy due to interparticle repulsion. 
Such a fragmented condensate can be regarded as having 
two entirely independent 
condensates.  If we changed the potential very
fast, then we would obtain a single condensate that oscillates
in each of the potential wells. We would also expect to see
collapses and revivals in the ``condensate phase'' \cite{prl77-2159}, provided the
losses are not important \cite{EPJD4-247}.

It is clear that the Gross--Pitaevskii Equation (GPE) will not
give a good description of the splitting process, in principle.
This equation describes the evolution of a single mode of the
condensate $\varphi(z,t)$, and, therefore, is not valid for
fragmented condensates. In order to interpolate between the 
Gross-Pitaevskii limit of phase coherence and the limit of
two independent condensates, one needs to consider at
least two modes $\varphi_{1,2}(z,t)$. Then, one can write
the state of the atoms as (we assume $N$ to be even)

\begin{eqnarray}\label{sumc_m}
|\Phi(t)\rangle&=&\sum_{m=-N/2}^{N/2} c_m(t) |m(t)\rangle,
\end{eqnarray}
where

\begin{eqnarray}
|m(t)\rangle&=&
\frac{a_1(t)^{\dag \frac{N}{2}-m}}{\sqrt{ \left( \frac{N}{2}-m \right) !}}
\frac{a_2(t)^{\dag \frac{N}{2}+m}}{\sqrt{ \left( \frac{N}{2}+m \right) !}} |vac 
\rangle,
\end{eqnarray}
and $a_{1,2}$ are the mode annihilation operators defined as

\begin{eqnarray}
a_i^{\dag}(t)=\int \varphi_i(z,t) \Psi^{\dag}(z) \, dz .
\end{eqnarray}
We have to consider the evolution of the
wavefunctions $\varphi_{1,2}$ as well as of the coefficients
$c_m$, which will be coupled and governed by the Hamiltonian

\begin{eqnarray} \label{2modeh}
&&\hat{H}(t)= \\
&&=\sum_{ij=1,2} a_i^{\dag} a_j   \int \varphi_i^*(z,t) 
\left( \frac{p^2}{2M} + V(z,t) \right) \varphi_j(z,t) dz   \nonumber   \\
&&+\frac{g}{2} 
\sum_{ijlm=1,2}  a_i^{\dag} a_j^{\dag}  a_l a_m  \nonumber  \\
&& \hspace{2cm}  \int 
\varphi_i^*(z,t) \varphi_j^*(z,t) \varphi_l(z,t) \varphi_m(z,t) dz. 
\nonumber 
\end{eqnarray}
We will refer to the evolution of those wavefunctions as
``external dynamics'' and to the one of the coefficients as
``internal dynamics''.

In order to find the mode--functions $\varphi_{1,2}$ we can use
the variational principle (in the same way as one derives the
GPE from a Hartree-Fock ansatz). 
However, this also turns out to be very complicated. A way
around that problem is to express $\varphi_{1,2}$ in terms of
some few variational parameters: we will use  quasi--gaussian 
functions to describe those mode-functions. On the
other hand, we could also use a variational principle to
determine the evolution of the coefficients $c_m$. This again
turns out to be very complicated, so that we will also use a
Gaussian ansatz for them.
Once they are known, 
in order to estimate when the condensate is fragmented, we can
look at the eigenvalues of the single particle density operator
corresponding to the internal dynamics only, that is, the matrix

\begin{eqnarray}
\rho= \frac{1}{N} \left( \begin{array}{c c}
\langle a_1^{\dag} a_1 \rangle & \langle a_1^{\dag} a_2 \rangle\\
\langle a_2^{\dag} a_1 \rangle & \langle a_2^{\dag} a_2 \rangle
\end{array}
\right),
\end{eqnarray}
where $ \langle a_i^{\dag} a_j \rangle $ mean the expectation
value on the state $|\Phi \rangle$.
The eigenvalues $\lambda_{\pm}$ indicate whether  we have a single
condensate ($\lambda_+ \sim 1$, $\lambda_- \sim 0$), or a
fragmented one ($\lambda_+ \sim \lambda_- \sim 1/2$).

As already extensively discussed in the literature
\cite{pra55-4318,pra60-2351}, the two-mode model has a limited
validity in the case of low barrier, when in principle one is
not allowed to neglect higher excited modes. However, it becomes
more and more accurate the higher the barrier gets, since in
this case the two lower lying modes move closer together in
energy compared to the higher ones. Hence it should allow a good
description of the splitting process.

\subsection{Qualitative behavior}

Before presenting the numerical and analytical result coming
from our analysis, we will briefly discuss the qualitative
behavior that we expect from the model under study. We will 
show  that the equations we will derive for the external and
internal dynamics can be decoupled to a very good approximation.
That is, we can first solve the equations for the external
dynamics basically taking constant values for the variational
parameters describing the coefficients $c_m$. Once these
equations are solved, we can use the corresponding wavefunctions
$\varphi_{1,2}$ to calculate the time--dependent coefficients
for the equations that describe the internal variational
parameters. In summary, once we have solved the equations for
the external parameters, we are left with a two--mode model with
time--dependent coefficients which contain all information
about the external dynamics: they will define the hopping and
on-site interaction, whose competition determines the phase
relation between the two modes.

Regarding the external dynamics, one can see two kinds of
behaviors depending on the time scale $\tau$ at which the
barrier is raised. The important time scale with which one has to
compare is the oscillation period $\tau_z$ in the trapping
potentials at each time. These periods change by roughly a
factor of two between the initial harmonic potential and the
final double well (with our specific choice of the trapping
potential). Thus, if 
$\tau \gg \tau_z$ the process will be adiabatic with 
respect to the external dynamics, which means that 
$\varphi_{1,2}(z,t)$ will basically correspond to the two
ground states of the right and left wells at the final time. If
$\tau \ll \tau_z$ we will have collective excitations, in which
$\varphi_{1,2}(z,t)$ oscillate strongly. In this case, we will
have that the energy of the condensate $E$ (with respect to the
ground state energy) has increased, so that it may be destroyed.
Although we cannot account for the disappearance of the
condensate within our model, we can estimate when this will
happen just by considering the fact that under normal
circumstances thermalization will occur, and, therefore, the
condensate will be destroyed for $E=K_BT_c>N \hbar\omega_z$, where
$\omega_z=2\pi/\tau_z$, $K_B$ is the Boltzman constant, $T_c$ indicates
the critical temperature and $E$ is the extra energy in the final 
state.
We find that the condensate disappears for $\tau\simeq \tau_z$.

Regarding the internal dynamics, there is also a time scale in
the problem which determines the dynamics; this is the revival
time $\tau_{r}$. Given two condensates with an initial  
well defined relative phase, it is well--known that
the relative phase first disappears (collapse) and then is restored 
at time $\tau_{r}\sim 1/g$ \cite{jourmodopt44-1775,EPJD8-319}. 
If $\tau\gg \tau_{r}$ the process will be adiabatic with respect to
the internal dynamics, the phase coherence will be lost during the 
process and therefore we will end up with two
independent condensates in each well, with no phase coherence at
all (note that it makes no sense to talk about collapses or
revivals in this situation). If $\tau\ll \tau_r$ at the end of 
the process we will have two condensates with a
good phase coherence. In that case, after the splitting,
collapses and revivals could be observed provided the particle
losses are practically absent.

In summary, we have two important time scales in the problem,
namely $\tau_z$ and $\tau_r$. Typically, in experiments
$\tau_r\gg \tau_z$, so that it will be harder to be adiabatic
with respect to the internal dynamics than to the external one.
On the other hand, since $\tau_r$ is very long in practise,
it will be hard to achieve $\tau\gg \tau_r$ within the validity 
of our model, in which particle losses and other imperfections 
are not included.

Finally we notice that the external and internal dynamics depend
in a very different way on the parameters $g$ and $N$.
Very similar to what happens in the GPE, the equations of motion
for the parameters describing the external dynamics
depend almost only on the product $gN$. On the contrary, the
on-site energy splitting scales like $g$ and the tunneling coupling
like $N$ giving rise to very different internal dynamics.
For increasing $N$ and decreasing $g$ the relative phase
becomes better defined and the time $\tau_r$ required to destroy
the phase coherence  is longer. In particular in the limit 
$N\rightarrow \infty$,  unless the tunneling coupling  exactly 
vanishes, the GPE case of phase coherent condensate is recovered.

\section{Variational ansatz}\label{ansatz}

In this Section we introduce a variational ansatz to
describe the internal and external dynamics. To describe the 
ground state of the system, which we will call equivalently static
or equilibrium solution, two parameters are sufficient: $z_0$ which
corresponds roughly to the center of the mode functions, and $p$ 
which is related to the width of the number distribution. To allow 
dynamic evolution we have to introduce also the variables 
$\sigma_I$ and $x_I$, which will vanish in the static case.
For simplicity in the following we present the variational ansatz
for the symmetric case, describing a spatially symmetric double--well 
potential and a symmetric atomic distribution among the two modes.
Starting from symmetric initial conditions
(no unbalance in the population of the two modes and symmetric
mode functions $\varphi_1(z)=\varphi_2(-z)$), 
the symmetry will be preserved under the time
evolution. Nevertheless, it is possible to generalize our
ansatz to describe asymmetric situations. 
We will briefly discuss the corresponding results in the conclusions.

\subsection{Variational ansatz for the mode functions}\label{ansatz1}

For a single condensate in a harmonic trap, a Gaussian ansatz
for the wavefunction has proved to be very useful and able to
predict the excitation frequencies within a very high precision
\cite{prl77-5320}.
In our case we make a similar choice. For a very high barrier we
expect to find two separate condensates for each of which a
Gaussian ansatz should be good. We then define the two functions
for the left and right side:

\begin{eqnarray}\label{variationalvarphi}
&& \varphi_{R,L}(z)=   \\ 
&=&  \left( \frac{\sigma_R}{\pi} \right)^{1/4} 
\exp\left[-\frac{\sigma_R (z\mp  z_0)^2}{2} \right]   
\exp\left(-i \frac{\sigma_I}{2}z^2 \right).   \nonumber
\end{eqnarray}
In the case of low barrier (small $z_0$), these two functions
have to be orthogonalized to satisfy the orthonormality property
required for the two mode functions. Hence we define

\begin{mathletters}
\begin{eqnarray}
\varphi_{\pm}(z) &=& \frac{1}{\sqrt{2}\sqrt{1\pm\exp(-\sigma_Rz_0^2)}} \left( 
\varphi_R \pm \varphi_L \right) , \\
\varphi_{1,2}(z) &=& \frac{1}{\sqrt{2}} \left( \varphi_+ \pm \varphi_- \right),
\end{eqnarray}
\end{mathletters}
which are different from $\varphi_{R,L}$ because of the two
different normalization constants for $\varphi_{+}$ and
$\varphi_{-}$.

The variational parameters describing the mode functions
$\varphi_{1,2}$ are $z_0$ and  $\sigma_I$.
Physically, the excitation modes that these parameters can
describe depend on whether the two Gaussians significantly
overlap in space or not. If they do, as in the case of low barrier,
then the excitation mode (changes in $z_0$) corresponds to a 
breathing mode. If they do not overlap, as for high barrier,
then it corresponds to an oscillation mode in each of
the potential wells. Allowing also $\sigma_R$ to vary (and adding
an extra variable for the dynamics) one can describe more
excitation modes. Instead we fix it to a constant value, since
the curvature of the potential wells will be chosen to be always
of the same order of magnitude. The value of $\sigma_R$ is not
obvious since the overlap integrals depend strongly on it. For
instance the hopping terms can be overestimated due to the long
Gaussian tales. To fix  $\sigma_R$, we identified a range of 
values for which the dynamic behaviour of the system 
was qualitatively the same and choose one of the values within
this interval, which turned out to be lower than the one corresponding 
to the static solution.

Since the mode wavefunctions are linear combinations of
Gaussians, if we choose a trapping potential of the form

\begin{eqnarray}\label{trappot}
V(z)=\frac{1}{2} M \omega_T ^2 z^2+ V_0(t) \exp[- z^2/a(t)] ,
\end{eqnarray}
the integrals in Eq.(\ref{overlapint}) can be performed
analytically.

In the following, we will use dimensionless units:
$a_{ho}=\sqrt{\hbar/M\omega_T}=1$, $\hbar \omega_T=1$ and  $\hbar=1$.
So, all lengths will be measured in units of harmonic oscillator length,
all energy in units of the trap frequency and all times in units
of $\omega_T^{-1}$.

\subsection{Variational ansatz for the coefficients}\label{ansatz2}

We also take for the coefficients $c_m(t)$ a Gaussian
distribution centered at $m=0$,
\begin{eqnarray}
c_m&=&{\cal N}(p)
\exp \left[- \left( \frac{1}{4 { p}}+i { x_I} \right)m^2 \right] ,
\label{coeff}
\end{eqnarray}
where ${\cal N}(p)$ is a normalization constant that depends on
$p$ only. The variational parameters are $p$ and its conjugate
one $x_I$. The parameter $p$ is directly related to the width of
the distribution $|c_m|^2$, whereas $x_I$ contributes to the
width of the Fourier transform of such a distribution (i.e., to
the width of the phase distribution, see App.~\ref{varanephmod}).

\subsection{Time-dependent variational principle}

We study the dynamics using the time dependent variational
principle. To derive the equations of motion one starts by
writing the action $S$

\begin{eqnarray}
S=\int dt \frac{\langle \dot{\Phi} | \Phi \rangle - \langle \Phi
| \dot{\Phi} \rangle }{2i} -\langle \Phi| \hat{H} | \Phi
\rangle .
\end{eqnarray}
where ${\hat H}$ and $|\Phi\rangle$ have been defined in (\ref{2modeh})
and (\ref{sumc_m}), respectively.
In evaluating the term $\langle \Phi | \dot{\Phi} \rangle$, one
should remember that the state $|\Phi\rangle$ depends on time
both through the coefficients and the mode functions contained
in $|m\rangle$:

\begin{eqnarray}
\langle {\dot \Phi} | \Phi \rangle = \sum_m {\dot c}_m^* c_m +
\sum_{mm'} c_{m'}^* c_m  \langle {\dot m}' | m \rangle.
\end{eqnarray}
The Lagrangian which follows takes the form

\begin{eqnarray}\label{lagr1}
L &=&\frac{1}{2i}\left[ \sum_m  {\dot c}_m^* c_m  + \right. \\  &&  \left. \hspace{1cm}
 +\sum_{ij} \langle a_i^{\dag} a_j \rangle 
\int {\dot  \varphi}_i^* \varphi_j dz - h.c. \right] \nonumber
- {\cal H} ,   
\end{eqnarray}
where ${\cal H}\equiv\langle \Phi | {\hat H} | \Phi \rangle$ is given by

\begin{eqnarray} \label{2modehexpval}
&&{\cal H}= \\
&&=\sum_{ij=1,2}  \langle a_i^{\dag} a_j \rangle   \int \varphi_i^*(z,t) 
\left( \frac{p^2}{2M} + V(z,t) \right) \varphi_j(z,t) dz  \nonumber    \\
&&+\frac{1}{2}\;g
\sum_{ijlm=1,2}  \langle a_i^{\dag} a_j^{\dag}  a_l a_m \rangle \nonumber \\
&& \hspace{2cm}
 \int \varphi_i^*(z,t) \varphi_j^*(z,t) \varphi_l(z,t) \varphi_m(z,t) dz.  \nonumber
\nonumber 
\end{eqnarray}
From the Lagrangian (\ref{lagr1}), in general one derives the equations of 
motion, carrying out the variation with respect to the discrete variables
$c_m$ and the fields $\varphi_i$. In our case, all the integrals and 
expectation  values are functions of the variational parameters which 
can be calculate analytically. The only important overlap integrals 
containing time derivatives are

\begin{mathletters}
\begin{eqnarray}
\int \dot{\varphi}_{1,2}^*\varphi_{1,2} dz &=&
-\frac{i}{2}
\left( \frac{1}{\sigma_R} +\frac{z_0^2}{1-e^{-2 \sigma_R z_0^2 } } \right) 
\dot{\sigma}_I , \\
\int \dot{\varphi}_{1,2}^*\varphi_{2,1} dz &=&
\frac{i}{2} \frac{z_0^2 e^{-2 \sigma_R z_0^2 } }{1-e^{-2 \sigma_R z_0^2 } }  
\dot{\sigma}_I .
\end{eqnarray}
\end{mathletters}
If one defines 

\begin{eqnarray}
N_{\phi}=\langle a_1^{\dag} a_2 \rangle +\langle a_2^{\dag} a_1 \rangle
\label{nphi}
\end{eqnarray}
and

\begin{eqnarray}
&&{\cal I}(p,x_I,z_0) \equiv \\
&=&\frac{N}{2} \left( \frac{1}{\sigma_R} +\frac{z_0^2}{1-e^{-2 \sigma_R z_0^2 } } 
\right)
-\frac{N_{\phi}}{2} \frac{z_0^2 e^{-2 \sigma_R z_0^2 } }{1-e^{-2
\sigma_R z_0^2 } } ,  \nonumber
\end{eqnarray}
the Lagrangian becomes

\begin{eqnarray}
L&=& p \dot{x}_I +{\cal I} \dot{\sigma}_I-
\cal{ H },
\end{eqnarray}
and the corresponding equations of motion are

\begin{eqnarray} \label{eqmotion}
&&\left( \begin{array}{c c c c}
0 & -1 & 0 & -\frac{\partial {\cal I}}{\partial p} \\ 1 & 0 & 0
& -\frac{\partial {\cal I}}{\partial x_I}
\\ 0 & 0 & 0 & -\frac{\partial {\cal I}}{\partial
z_0} \\
\frac{\partial {\cal I}}{\partial p}  &
\frac{\partial {\cal I}}{\partial x_I}  &
\frac{\partial {\cal I}}{\partial z_0}  & 0
\end{array} \right)
\left( \begin{array}{c}
\dot{p} \\
\dot{x}_I \\
\dot{z}_0 \\
\dot{\sigma}_I
\end{array} \right) =
-\left( \begin{array}{c}
\frac{\partial {\cal H}}{\partial p}  \\
\frac{\partial {\cal H}}{\partial x_I} \\
\frac{\partial {\cal H}}{\partial z_0} \\
\frac{\partial {\cal H}}{\partial \sigma_I}
\end{array} \right)   \nonumber   \\
&\Rightarrow&
\left\{
\begin{array}{r c l}
\dot{p} &=&  - \frac{\partial {\cal H}}{\partial x_I}
                + \frac{\partial {\cal I}}{\partial x_I} \dot{\sigma}_I  \\
-\dot{x}_I &=& - \frac{\partial {\cal H}}{\partial p}
                + \frac{\partial {\cal I}}{\partial p} \dot{\sigma}_I \\

\frac{\partial {\cal I}}{\partial z_0} \dot{z}_0  &=&
 - \frac{\partial {\cal H}}{\partial \sigma_I}
                - \left(  \frac{\partial {\cal I}}{\partial p} \dot{p} +
                         \frac{\partial {\cal I}}{\partial x_I} \dot{x_I} 
\right) \\
-\frac{\partial {\cal I}}{\partial z_0}
\dot{\sigma}_I &=&
 - \frac{\partial {\cal H}}{\partial z_0}
\end{array}
\right.
\end{eqnarray}
These equations describe the internal and external coupled dynamics of
the splitting of the condensate. In what follows, we have solved
them numerically in different regimes. Before presenting the
results we will show that one can decouple the evolution of the 
external and internal variables, which helps  to understand the dynamics.
For that, we will introduce in the next subsection
some analytical approximations to derive explicit formulae for
the quantities ${\cal H}$ and ${\cal I}$ by replacing the
discrete distribution $c_m$ by a continuous one, and treating the
index $m$ as a continuous variable running from $-\infty$ to
$\infty$.

\subsection{Continuous limit}

In order to calculate ${\cal H}$ and ${\cal I}$ we have to
evaluate expectation values of the form $\langle a_i^\dag a_j \rangle$ 
and $\langle a_i^\dagger a_j^\dag a_l a_m \rangle$. 
We can do that if we replace the sums in
$m$ by integrals extended from $-\infty$ to $\infty$. When this
replacement is valid, we can even calculate the width of the
number distribution $|c_m|^2$, $\sigma_m$, as well as the one
corresponding to the phase distribution, $\sigma_\phi$ (see App.\ref{varanephmod}). 
We obtain

\begin{mathletters} \label{sigmas}
\begin{eqnarray}
\sigma_{m}&=&\sqrt{p},  \\
\sigma_{\phi}&=&
\sqrt{\frac{1}{4p}+4px_I^2}.  \label{sigmasphi}
\end{eqnarray}
\end{mathletters}
On the other hand, we have

\begin{mathletters} \label{expaaaa}
\begin{eqnarray}
\label{expaaaan1}
&& \langle  a_{1,2}^{\dag} a_{1,2}  \rangle =
\frac{N}{2} , \\
\label{expaaaan2}
&& \langle  a_{1,2}^{\dag2} a_{1,2}^2  \rangle =
\frac{N^2}{4}   + p , \\
\label{expvalaa}
&& \langle  a_1^{\dag} a_2 \rangle =
\langle  a_2^{\dag} a_1  \rangle^* =  \\
&& \hspace{0.5cm}= \frac{N}{2} \exp \left(-\frac{\sigma_{\phi}^2}{2}\right)
\left[1-\frac{2p}{N^2} +\frac{8p^2x_I^2 }{N^2} \right],   \nonumber       \\
&& \langle  a_1^{\dag} a_2^{\dag} a_1 a_2  \rangle =
\frac{N^2}{4} -p ,\\
\label{expvalaa2}
&& \langle  a_1^{\dag2} a_2^2  \rangle = \\
&&
\hspace{0.5cm} = \frac{N^2}{4}  \exp \left( -2\sigma_{\phi}^2 \right)
  \left[ 1 -  \frac{4 p}{N^2}   +  \frac{64 p^2 x_I^2 }{N^2}  \right] .   \nonumber
\end{eqnarray}
\end{mathletters}
Furthermore, in this limit we can determine the eigenvalues of
the single particle density operator corresponding to the
internal degrees of freedom, obtaining

\begin{eqnarray}
\lambda_{\pm} \approx
\frac{1 \pm \exp \left(-\sigma_{\phi}^2/ 2 \right) }{2}.
\end{eqnarray}
Notice that $\lambda_+\rightarrow 1$ for $p\sim N/4,x_I\sim 0$, 
i.e. $\sigma_{\phi}  \sim 0$, which corresponds to the
Gross-Pitaevskii limit; instead $\lambda_{\pm}\rightarrow 1/2$ for
large $\sigma_{\phi}$, giving a signature of the fragmentation
of the condensate.

One can easily determine the limits of validity of this
continuous approximation. On the one hand, the distribution in
$c_m$ has to be sufficiently broad, which implies $p\agt 1$. On
the other hand, $x_I$ has to be such that $\sigma_\phi \alt
\pi$. 
For our numerical simulations, we corrected the expressions in Eqs.(\ref{expaaaa})
to make them valid $\forall p$ and $\forall x_I$: for $p>1$  we
included the periodicity in $x_I$ and for $p<1$ we performed the exact 
sum over $m$, considering only the few number states which are populated.

\subsection{Decoupling between external and internal dynamics}\label{decouple}

The coupling among the $z_0$ dynamics and the $p$ dynamics
appears in the off-diagonal blocks in Eq.(\ref{eqmotion}) and in
the dependence of ${\cal H}$ on all variational parameters. In
the following we will analyze under which condition it is
possible to decouple the internal ($p,x_I$) and external
($z_0,\sigma_I$) dynamics, so that one can study them
independently one from the other.

\subsubsection{External dynamics}

One can rewrite the equation of motion for $z_0$ is a more
explicit form as

\begin{eqnarray}
\frac{\partial {\cal I} }{\partial z_0} \dot{z}_0 &=&
-\frac{\partial \cal{H}}{\partial \sigma_I} +
\frac{1}{2} \frac{z_0^2 \exp(-\sigma_R z_0^2)}{1-\exp(-2\sigma_Rz_0^2)} 
\dot{N}_{\phi} .
\end{eqnarray}
Let us first see under which circumstances it is possible to
neglect the off-diagonal blocks:

(i) in the low barrier limit, i.e. $\exp(-\sigma_R z_0^2) \sim 1$,
the off-diagonal blocks can be neglected if ${\dot N}_{\phi}$
plays no role:  $N_{\phi}$ evolves at the frequency which governs
the internal dynamics $\omega_p$ (remember definition (\ref{nphi})), 
while the external dynamics is governed by the frequency $\omega_z$,
usually of the order of the trapping frequency $\omega_T$.
If $\omega_p\gg \omega_z$, then the time average 
of $\dot{N}_{\phi}$ vanishes and has no effect
on the evolution of $z_0$. On the other hand
$\omega_p \sim gNU_1 \exp(-\sigma_R z_0^2/2)  \alt \omega_z$
only for a chemical potential $\mu \sim gNU_1 \alt \omega_T$.
This happens either for very few particles, a case in which our 
model does not hold, since $N \gg 1$ is a fundamental assumption
in our model, or for very weakly interacting particles, 
in which ${\dot N}_{\phi} \sim 0$, because the phase coherence is very
difficult to be destroyed.

(ii) in the high barrier limit, normally $\omega_p \ll \omega_z$ 
(see Fig.~\ref{posc}, small $p$ limit) and $N_{\phi}$ might vary abruptly, 
since this is  when the phase coherence is supposed to desappear.
Anyway, since $\exp(-\sigma_R z_0^2) \sim 0 $, the
off-diagonal block can be neglected.

Now let us analyze the $p$--$x_I$ dependence in ${\cal H}$
(see Eqs.(\ref{2modehexpval},\ref{expaaaa})). In the on-site terms 
this dependence is of order  $p/N^2, x_I/N^2 \ll 1$ and can be therefore 
safely  neglected (Eqs.(\ref{expaaaan1},\ref{expaaaan2})). 
In the hopping terms, it scales like  $\exp(-\sigma_{\phi}^2/\alpha)$ 
($\alpha= 2,1/2$, see Eqs.(\ref{expvalaa},\ref{expvalaa2})): it is 
strong only when the hopping terms are already small and negligible 
in comparison with the on-site terms, so it can also be neglected.

\noindent
After these considerations, we conclude that the $z_0$-dynamics
is, in a good approximation and in reasonable regimes,
independent of the $p$-dynamics. This is confirmed by the numerical
simulations.

\subsubsection{Internal dynamics}

The off-diagonal blocks can be neglected in the $p$-dynamics if
$\partial {\cal H} / \partial z_0 \sim 0$ or 
$\exp( -\sigma_R z_0^2 ) \sim 0$.
If the barrier is raised starting from a condensate in the ground state,
$z_0$ evolves almost adiabatically in the low barrier limit
($\partial {\cal H} / \partial z_0 \sim 0$); when it reaches the
high barrier limit, then $\exp( -\sigma_R z_0^2 ) \sim 0$.
Therefore, the off-diagonal terms can be neglected during the
whole process.

Instead, the $z_0-\sigma_{I}$ dependence in ${\cal H}$ is
strong. One can solve the complete coupled dynamics or
substitute in ${\cal H}$ the adiabatic solution for $z_0$ and
compare the results. We will show some examples in the following
and see that the difference is small.

\section{External dynamics: excitations}\label{finexc}

In the previous section, we have written the equations of motion
describing the full coupled dynamics and demonstrated that
if one splits a condensate starting from a ground state configuration,
it is possible to decouple the internal and the external dynamics.
In this section, we will use this result and study the external
dynamics decoupling it from the internal one. 
Making  use of the external static solution, in Sec.~\ref{numres} 
we will discuss the static solution for the internal degrees of freedom.
Finally in Sec.~\ref{results}, we show the results for the internal
dynamics, which are also useful as a check of the decoupling assumption.

The fact that the internal and external dynamics decouple 
under the condition discussed above,
allows us to estimate the excitation of the collective modes
using a very simple model. We point out that in
this case, similarly to the Gross-Pitaesvkii equation, the external 
dynamics is governed by the product $gN$ and not by the two quantities
separately.

We choose to raise the potential barrier with the following time dependence

\begin{eqnarray}
V_0(t)=\frac{V_{0fin}}{2} \left( \tanh \frac{t}{\tau} +1 \right).
\end{eqnarray}
From the static solution, obtained by minimizing ${\cal H}(p,x_I,z_0,\sigma_I,V_0)$
with respect to all variational parameters simultaneously at fixed $V_0$, 
we know the equilibrium position of $z_0$ at any value
of the barrier. The $z_0$ dynamics can be approximated very well by the dynamics
in a harmonic potential whose center
 moves from $z_{0in}$ to $z_{0fin}$ following the adiabatic solution and whose 
frequency
changes corresponding to the frequency of the small oscillations.

To get analytic estimations, we model this dynamics  by  fixing the frequency 
and
shifting  the center of the  potential along a   trajectory  $z_c(t)$ for which 
an
analytic solution exists.
When  the center follows an  hyperbolic
tangent trajectory with time constant $\tau$, the semi-amplitude of the 
oscillation is given by

\begin{eqnarray}
&&\delta z_0 = \\
&& \left( z_{0fin}-z_{0in} \right) \frac{\pi \omega \tau}{4}
{\rm csch} \left( \frac{\pi \omega \tau}{4} \right)
{\rm sech} \left( \frac{\pi \omega \tau}{4} \right).   \nonumber
\end{eqnarray}
Otherwise, if the center moves according to a linear ramp with time constant 
$\tau$, the average semi-amplitude of the oscillation is

\begin{eqnarray}
\delta z_0 &=& \sqrt{2} \frac{z_{0fin}-z_{0in}}{\omega \tau}.
\end{eqnarray}

\begin{figure}[htb]
\begin{center}
\epsfig{file=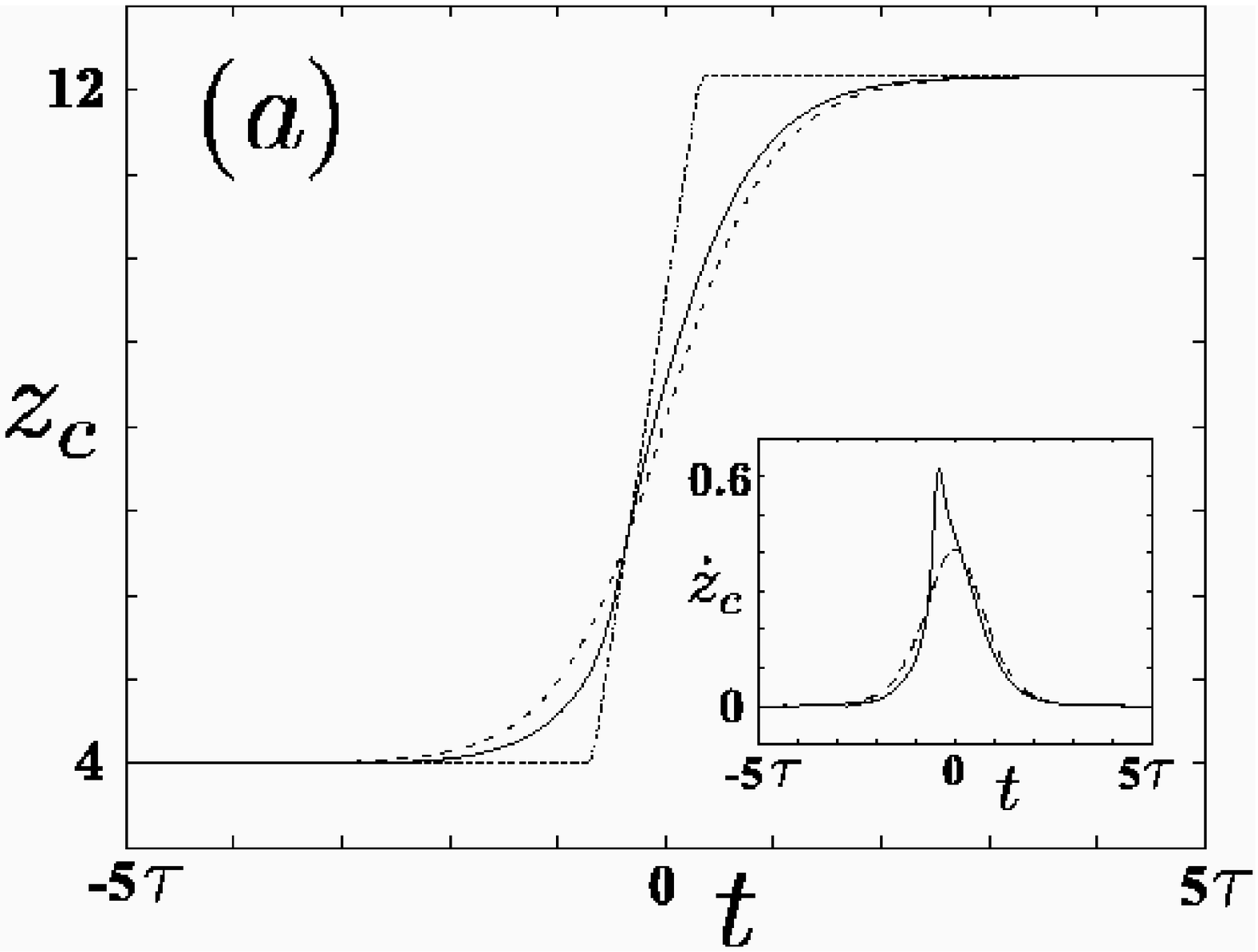,height=0.33\linewidth}
\epsfig{file=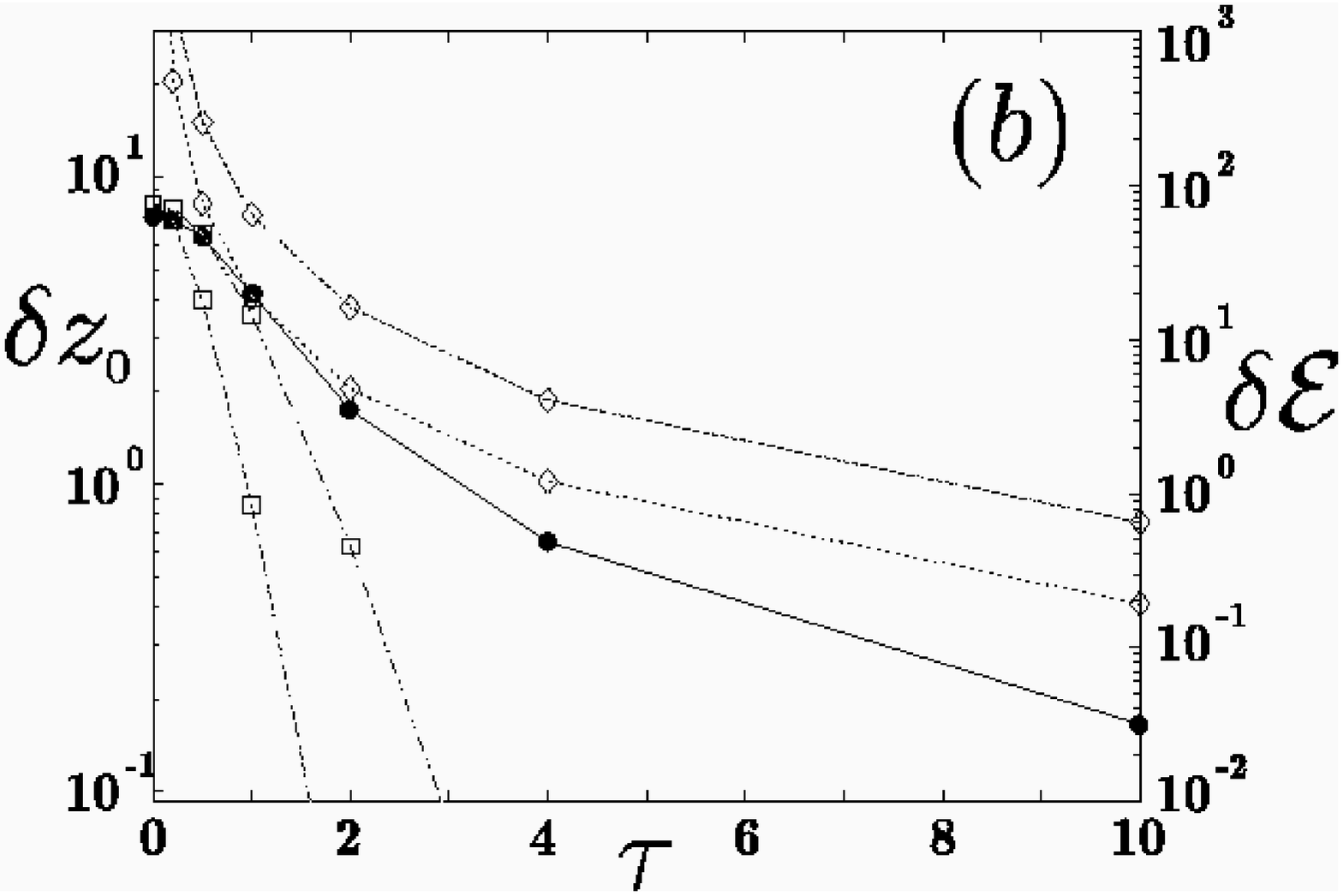,height=0.34\linewidth}
\end{center}
\caption{(a) Static solution for $z_0$ (full line) and position for the center of
the harmonic potential for a raising time scale $\tau$: hyperbolic tangent shift 
(dashed line)
and linear shift (dotted line); time derivative of $z_0(t)$ (inset).
(b) Semi-amplitude of the final oscillations (left axis) and final excitation 
energy per particle (right axis): numerical results ($\bullet$), results
for the $\tanh$ ($\Box$) and linear shift ($\diamond$) for $\omega=2.82$ and 
$1.54$.}
\label{z0osc}
\end{figure}
We compare in Fig.\ref{z0osc} these expressions with the numerical results.
For fast raising
of the barrier the hyperbolic tangent
shift gives a good estimate. For slower raising, instead, the
amplitude of the oscillations is largely underestimated. In this case, the 
linear shift of the
center is useful to give an upper bound. We have checked numerically (by 
changing the
frequency of the harmonic potential in time according to the adiabatic
solution)
that the small discrepancy between the actual
shift of the center and the hyperbolic tangent dependence is enough to produce 
such
a big change in the amplitude of the final oscillations.
The reason for this might be better understood comparing the time derivative
of $z_c(t)$ in the two cases (see inset in Fig.\ref{z0osc}(a)). Our estimations 
are
qualitative, but allow us to set some lower and upper bounds and deduce an 
adiabaticity
condition $\tau \ll 1/\omega_z$ for the external degrees of freedom.
In the same way we can estimate the extra energy per particle due to the
excitation of collective modes. Using the relation $K_BT_c \sim N \hbar 
\omega_z$
(the trapping frequency at the end of the process is $\omega_z$ and not
$\omega_T$), we can estimate which is the fastest time scale which does not 
destroy
the condensate. As expected, it is $\tau \sim 1/\omega_z$.

\section{Internal dynamics: static solution}\label{numres}

In this section, we will study the complete splitting 
process, starting from a condensate trapped in a harmonic
potential and ending with a potential barrier with height
$V_{0}$.
Apart from presenting the numerical results obtained by solving
Eqs.(\ref{eqmotion}), we will also introduce a simple two--mode model 
that allows a deeper understanding of the results obtained by
the variational ansatz.

\subsection{Two--mode model}

Given the fact that the external dynamics is basically
independent of the internal one, we can derive a simple model
that accounts for most of the effects related to the internal
dynamics. The Hamiltonian in Eq.(\ref{2modeh}) depends on the 
following overlap integrals

\begin{mathletters} \label{overlapint}
\begin{eqnarray}
K_{ij}&=&\int \varphi_i^*(z)
\frac{p^2}{2M} \varphi_j(z) dz , \hspace{0.5cm}  \\
V_{ij}&=&\int \varphi_i^*(z) V(z) \varphi_j(z) dz , \\
U_{1}&=&
\int |\varphi_i(z)|^4 dz , \\
\label{u2}
U_2&=&
\int |\varphi_i(z)|^2 |\varphi_j(z)|^2 dz =  \\
&=&
\int \varphi_i^*(z) \varphi_i^*(z) \varphi_j(z) \varphi_j(z) dz ,  \nonumber \\ 
\label{u3}
U_{3}&=&
\int |\varphi_i(z)|^2 \varphi_i^*(z) \varphi_j(z) dz ,
\end{eqnarray}
\end{mathletters}
where $i,j=1,2$ \cite{overlap}.
We use
\begin{eqnarray}
&& a_1^{\dag} a_1^{\dag} a_1 a_2  +  a_1^{\dag} a_2^{\dag} a_2 a_2  = \\
&&=
\left( a_1^{\dag} a_1 -1 + a_2^{\dag}  a_2  \right)  a_1^{\dag} a_2 
\approx   N a_1^{\dag} a_2 ,\nonumber 
\end{eqnarray}
to define two effective single-particle hopping terms
$J_{12}=-K_{12}-V_{12} - g N U_3=-K_{21}-V_{21} - g N U_3$.
The static solution for the external dynamics is known from
the minimization of ${\cal H}(p,x_I,z_0,\sigma_I,V_0)$. Plugging
the corresponding solution $z_0(V_0)$ in Eqs.(\ref{overlapint}),
we get Hamiltonian parameters depending only on the barrier
height $V_0$ (in particular $J_{12}=J_{21}=J$). Neglecting
constant terms, we write the simplified Hamiltonian

\begin{eqnarray} \label{expham}
{\hat H} &=&\frac{1}{2} g U_1 \left[ a_{1}^{\dag2} a_{1}^2 +
a_{2}^{\dag2} a_{2}^2\right] -J \left[ a_1^{\dag} a_2 + 
a_2^{\dag} a_1 \right] + \nonumber  \\
&+& 2 gU_2 a_1^\dag a_2^\dag a_1 a_2 + 
\frac{1}{2} gU_2 \left[ a_1^{\dag 2} a_2^2 + a_2^{\dag 2} a_1^2  \right] .
\end{eqnarray}
As it is well--known (see App.~\ref{varanephmod}), under certain conditions
we can replace this model Hamiltonian by a phase model of the form

\begin{eqnarray}\label{phmodelham2}
{\hat H}&=&-g \left( U_1-2U_2 \right) \frac{\partial^2}{\partial \phi^2}
-JN \cos \phi  + \\
&&+ g \frac{N^2}{4} U_2 \cos 2\phi \,,  \nonumber
\end{eqnarray}
where $\phi$ represents the relative phase between the two
modes. The overlap integral $U_2$ may be non-negligible at the beginning
of the process when the two mode functions overlap in a sensible way.
Instead at the end, when the two condensates are almost spatially separated
it is very small. In this case $U_2\rightarrow 0$ and one recoveres
the Josephson's Hamiltonian ~\cite{foundphys21-353}.
The ground state of such Hamiltonian is a localised
wavefunction for $JN \gg gU_1$ (corresponding to a broad number distribution)
and a delocalised  one for $JN \ll gU_1$ (corresponding to 
a narrow number distribution).

\subsection{Static solution and check of the Gaussian ansatz}\label{statsol}

For this simplified two--mode model, we write analytic
approximated espressions for the static solution and check the
validity of the Gaussian ansatz. As explained above, we let the parameters
$J$, $U_1$, $U_2$ depend on the barrier height $V_0$, according to
the static solution. 
The expectation value of the Hamiltonian can be now written as a function
of the internal degrees of freedom only, ${\cal H}(p,x_I,V_0)$, and  
allows to study the coherence properties of the ground state at
the different stages of the splitting process: for increasing
barrier height, we expect the fluctuations in the number
distribution to become smaller. 

\begin{figure}[htb]
\begin{center}
\epsfig{file=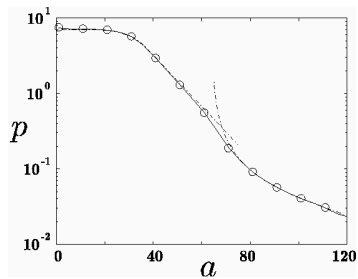,width=0.6\linewidth}
\end{center}
\caption{Static variational solution for $p$ as a function of $a$ (full line), 
analytic
solutions in the large and small $p$ limit (dashed-dotted line),
solution of the minimization of the exact Hamiltonian (circles).
All results are for $N=200$ and $g=5$; $a$ parametrize the
barrier (see Eq.(\ref{trappot}) with $V_0=ab$ \protect\cite{b}).}
\label{pstat}
\end{figure}

We solved the static problem numerically, finding the minimum of ${\cal H}$ 
with respect to $p$ and $x_I$ for fixed $V_0$. 
Moreover, in  the limits of large $p$ ($\exp(-\sigma_{\phi}^2/2) \sim 1$)
and small $p$ ($p\rightarrow 0$) it is possible to get analytic
estimation for the value of $p$ at equilibrium and the frequency
$\omega_p$ of the small oscillations as a function of the
overlap integrals \cite{jourmodopt44-1775,pra60-2351}.
In the large $p$ limit, one finds

\begin{mathletters}\label{largep}
\begin{eqnarray}\label{statplp}
&&p_s = \frac{N}{4} \sqrt{ \frac{2 (J-gNU_2) }
{  2 J-gNU_2  +gN(U_1-2U_2) }}, \\
&&\omega_p = \\ 
&&2 \sqrt{2 (J-gNU_2)[   2 J-gNU_2  +gN(U_1- 2 U_2) ]};  \nonumber 
\label{omegaplp}
\end{eqnarray}
\end{mathletters}
in the small-$p$ limit ($ JN \ll gU $), where the continuum approximation 
for $m$ is not valid, $p_s$ and $\omega_p$ can be calculated with perturbation
theory, considering the Hamiltonian in Eq.(\ref{expham}) with $J=U_2=0$
as unperturbed Hamiltonian and the number state $|m=0\rangle$ as unperturbed
ground state. Then we get

\begin{mathletters}\label{smallp}
\begin{eqnarray}
p_{s} &=& \left(4 \log \frac{2gU_1}{JN}\right)^{-1} , \\
\omega_{p} &=& gU_1.   \label{omegapsp}
\end{eqnarray}
\end{mathletters}
In Fig.~\ref{pstat} we plot the two analytic solutions for
$p_s$, comparing them with the numeric solution for the minimum
of $\cal{H}$ using our ansatz and the corresponding value of $p$
obtained by minimizing the exact Hamiltonian for $N=200$ and
$g=5$. The agreement between the variational solution and exact
one is excellent and the analytic expressions interpolate
correctly in the limits where they are expected to work.

Given the analytic espressions in Eqs.(\ref{largep},\ref{smallp}),
we observe that the oscillation frequency for large $p$
coincides with the breathing frequency in the harmonic potential
approximating the cosine potential, and the oscillation
frequency for small $p$ corresponds to the revival time when the
cosine potential is negligible (see Sec.~\ref{completesplit}). 
Aware of the fact that it is not
possible to define a transition point between the two regimes,
being it a smooth transition, we still calculate the value of $p$
at which the two frequencies coincide, get $p=0.125$ and claim
that the phase relation between the two condensates is
smeared out for $p<0.125$.

\section{Results: different regimes}\label{results}

In this section, we show some numerical results obtained by integrating
the equations of motion for the variational parameters. In those results
the {\it  full coupled dynamics} of the process were considered. 
It is possible to compare with the evolution of the phase 
distribution governed by the Hamiltonian (\ref{phmodelham2})
with time--dependent coefficients.
Moreover, it is possible to get analytic estimates concerning the
typical time scales of the process.

In the three cases presented below we will fix the product 
$gN$ in order to have similar external dynamics and better
isolate the effect of different $g$ and $N$ on the phase
properties of the system. At the end of the process, when 
the barrier has reached its final value, depending on $N$
and $g$, one can have $gU_1 \gg JN $ or  $JN \gg gU_1$ 
(we assume that $U_2$ is then negligible).
The case  $gU_1 \gg JN $ corresponds to the situation
where the splitting process is completed and one expects a ground
state with no well--defined relative phase.
Instead in the case $JN \gg gU_1 $ the ground state is still
characterized by a localized phase distribution, and even if the two
condensates are almost spatially separated they cannot be considered
as independent. In this sense, the splitting is not complete.

\subsection{Complete splitting}\label{completesplit}

We first analyse the case where in the final stage of the process,
one has $gU_1 \gg JN $. Since in this subsection and
in the following we fix the product $gN$, this case
is obtained for relatively small $N$ and large $g$.
Depending on the time scale of the process, it is possible to 
observe collapses and revivals or to reach a final fragmented condensate
characterized by very small number fluctuations.

\begin{figure}[htb]
\begin{center}
\epsfig{file=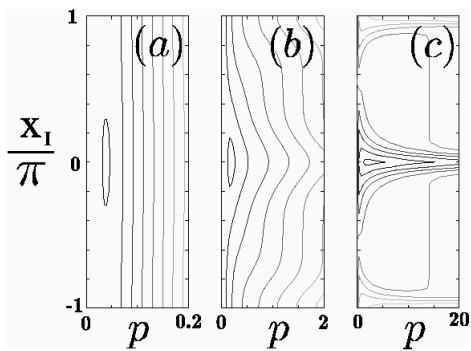,width=0.8\linewidth}
\end{center}
\caption{Contour plot of ${\cal H}$ for $a=120$: (a) $N=2 \times 10^3$, $g=0.5$, (b)
$N=2 \times 10^4$, $g=0.05$ and (c) $N=2 \times 10^5$, $g=0.005$ .}
\label{cplot}
\end{figure}

We assume that in the final stage of the process it is possible to 
neglect $JN$ and $N^2U_2$, since they depend on the overlap of the
two mode functions. Then the eingenstates are the number states 

\begin{eqnarray}
{\hat H} |m\rangle = g U_1 m^2 |m \rangle.
\end{eqnarray}
and the time evolution of the  final state corresponds to 
\begin{eqnarray}\label{phmodev}
&&|\Phi(t) \rangle \propto  \\
&&\sum_m
\exp \left( -\frac{ m^2}{4 p_{fin}} \right) \exp \left( -i g U_1 m^2 t \right) 
|m \rangle.  \nonumber
\end{eqnarray}
In the variational ansatz formulation, one can plot the constant energy 
trajectories 
for the final barrier height (see Fig.~\ref{cplot}(a)). Then, the time evolution 
corresponds to $p$ and $x_I$ following one of these trajectories: $p$ keeps 
almost constant and $x_I$ evolves unbounded increasing linearly with time 
with ``velocity'' $gU_1$ (see Eq.(\ref{omegapsp})), which
exactly reproduces the phase in Eq.(\ref{phmodev}). 

The main features of this time evolution are the following: the
width of the number distribution is constant (determined by $p_{fin}$) 
and do not evolve in time; instead the phase distribution collapses and 
revives: an initial distribution peaked around zero smears out, revives
around $\phi=\pi$ and so on.
Defining the collapse time as the time $\tau_c$ when
$\sigma_{\phi} \sim \pi $ and the revival time as the time
$\tau_r$ when the original phase distribution is recovered
shifted by $\pi$, one gets \cite{jourmodopt44-1775,pra55-4318}.

\begin{mathletters}\label{collrev}
\begin{eqnarray}
\tau_c &\sim& \frac{2}{4 p_{fin} g U_1} \\
\tau_r &=& \frac{\pi}{gU_1}.
\end{eqnarray}
\end{mathletters}
The collapse time is governed by $p_{fin}$, which
depends in general on the barrier raising process and on the
parameters $g$ and $N$.

The final width of the number  distribution obtained after the raising of 
the barrier is completed depends on the time scale of the process.
This is just given by the  value $p_{fin}$
at which the number fluctuations are frozen. To evaluated it,
we claim that as long as $\omega_p > 2\pi/\tau$ the number fluctuations
follow the static solution, and  when $\omega_p = 2\pi/\tau$,
they  are frozen out to a final value $p_{fin}$.
Setting $\omega_p = 2\pi/\tau$ in Eq.(\ref{omegaplp}) and substituting in
Eq.(\ref{statplp}), one gets \cite{jav}

\begin{eqnarray}\label{pfin}
p_{fin}&=&\frac{N}{4} \frac{1}{2[ 2 J-gNU_2 +gN(U_1-2 U_2)]} \frac{2\pi}{ \tau} 
\sim   \nonumber   \\
&\sim&\frac{1}{8gU_1}\frac{2 \pi }{ \tau}.
\end{eqnarray}
Of course Eq.(\ref{pfin}) is not valid for $\tau \rightarrow \infty$.
For $\tau > 2\pi/\ g U_1 = 2 \tau_r$, we are in the adiabatic regime:
$\omega_p > 2\pi/\tau$ during all the process, and we expect to reach a
completely delocalised relative phase.

\begin{figure}[htb]
\begin{center}
\epsfig{file=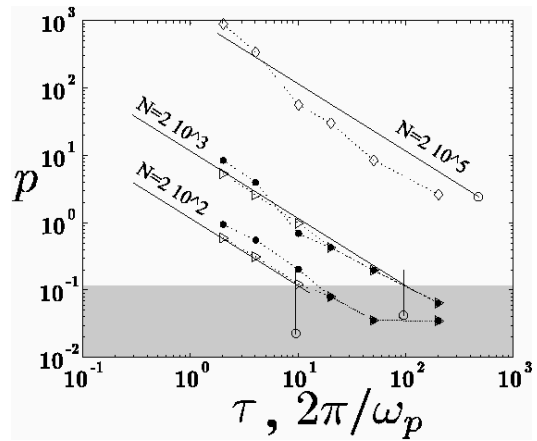,width=0.9\linewidth}
\end{center}
\caption{Case (i): $N=2 \times 10^2$, $g=5$ and  $N=2 \times 10^3$, $g=0.5$. Final
$p$ value against $\tau$:  complete dynamics ($\bullet$) and adiabatic external
dynamics ($\triangleright$); analytic solution for $p_{fin}$ against  
$2\pi/\omega_p$
in the large and small $p$ regimes (full lines);
the shaded area is for $p \leq 0.125$ and represent the situation of completely
delocalised relative phase;
Case (ii): $N=2 \times 10^5$, $g=0.005$;
analytic solution for $p_{fin}$ (full lines) and numerical value for $p_M$ 
($\diamond$);
Static solutions for $a=120$ ($\circ$) \protect\cite{b}).}
\label{posc}
\end{figure}

We check this  numerically and plot the results in Fig.~\ref{posc}
for several different splitting processes ($N=2 \times 10^2$ and $N=2 \times 10^3$). We
actually find that for time scales $\tau > 2 \tau_r$ the final $p$
values lie in the region $p<0.125$. For faster time scales, the estimation in 
Eq.(\ref{pfin})
was shown to be very good  when the external degrees of freedom evolve
adiabatically. Otherwise  discrepancies can be observed (Fig.~\ref{posc}).
It is not straighforward to estimate such discrepancies, since they depend on 
the
exact dynamics and can be either positive or negative. Anyway,
they are not striking and do not change from a  situation in which the final
relative phase is well defined to the opposite one.

\subsubsection{Collapses and revivals of the phase}\label{result1}

Now we consider two of these splitting processes more in detail and compare
quantitatively with the evolution of the phase distribution following
the Hamiltonian in Eq.(\ref{phmodelham2}), where $J$, $U_1$ and $U_2$ vary 
in time with the barrier height according to the instantaneous static solution. 

\begin{figure}[htb]
\begin{center}
\epsfig{file=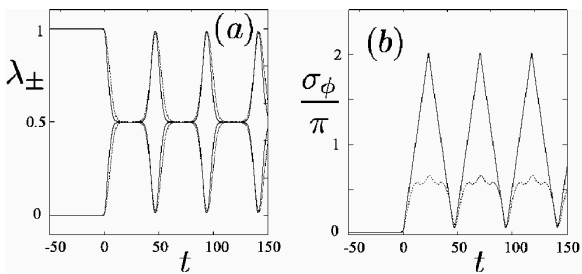,width=1\linewidth}
\end{center}
\caption{Eigenvalues $\lambda_{\pm}$ (a) and $\sigma_{\phi}/\pi$ (b). Numerical 
solutions
for the variational ansatz (full line) and for the phase model (dotted line);
$N=2 \times 10^3$, $g=0.5$ and $\tau=4$.}
\label{dyn1}
\end{figure}

We have already mentioned that collapse and revival of the phase
are predicted for two condensates with an initially good phase relation
when the final tunneling coupling is negligible.
Let us consider the case $N=2 \times 10^3$, $g=0.5$ and $\tau=4$. In Fig.~\ref{dyn1}
we show the one-atom density matrix eigenvalues and the indetermination in the
phase distribution $\sigma_{\phi}$.
The agreement between the results of the two simulations is perfect 
\cite{sigmaphiref} 
and  our analytic estimations  are also confirmed:
we expect $\tau_c \sim 8$ and $\tau_r \sim 50$,
which agree very well with  the numerical results shown in  Fig.~\ref{dyn1}. 

The actual possibility of observing the revivals of the phase in an experiment
is something outside our model. 
If they are actually destroyed by particles losses  \cite{EPJD4-247},
one is left  with two condensate with no phase relation, but higher number 
fluctations than they would have in the ground state.

\subsubsection{Final fragmented condensate}\label{result2}

Another way to cut the initial condensate into two independent ones, is to raise 
the
barrier much slower, so that the phase coherence is lost adiabatically all along 
the
process.
So, we now choose  again $N=2 \times 10^3$, $g=0.5$ but a longer time scale 
$\tau=200$.

\begin{figure}[htb]
\begin{center}
\epsfig{file=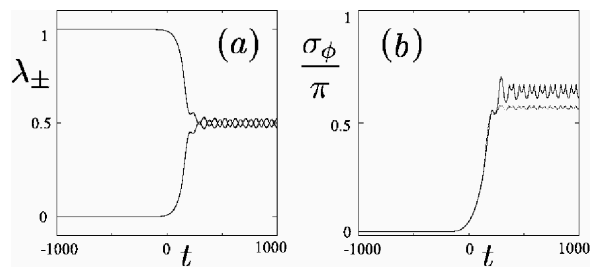,width=1\linewidth}
\end{center}
\caption{Eigenvalues $\lambda_{\pm}$ (a) and $\sigma_{\phi}/\pi$ (b). Numerical 
solutions
for the variational ansatz (full line) and for the phase model (dotted line);
$N=2 \times 10^3$, $g=0.5$ and $\tau=200$.}
\label{dyn2}
\end{figure}

The agreement between the results of the variational ansatz and the phase model
is very good also in this case. 
The final state is characterised by much smaller number fluctations
with respect to the previous case (see Fig.~\ref{dyn2}(a)) and
by a complete delocalised relative phase as shown in Fig.~\ref{dyn2}(b).

From our analytic estimations, we actually expect to reach the static
solution for $\tau>2\tau_r \sim 100$. Instead as it can be seen in 
Fig.~\ref{posc} for $N=2 \times 10^3,\tau=200$, this does not happen. 
Dealing 
with such small values of $p_{fin}$ it is very easy to get at the end very small
excitations, which can be due both to the external degrees of freedom or to
some excitations already present in the initial conditions. The important
feature  is that the final relative phase is anyway completely delocalised.

\subsection{Incomplete splitting}\label{result3}

We now analyse the case where in the final stage of the process,
one has $JN \gg gU_1 $.  This case is obtained for large $N$ and 
small  $g$: in the limit of large $N$, even if $J$
might be very small, it can happen that $JN>gU_1$ and the cosine
potential in the phase representation (Eq.(\ref{phmodelham2}))
is not negligible. This can be seen as a process in which the
barrier is raised up to a level at which the splitting is not really
completed.

In  the case in which the cosine potential at the end of the raising process 
is still deep, so that the lowest levels can be
approximated with harmonic oscillator levels spaced by
$\sqrt{2JNgU_1}$, the time evolution follows

\begin{eqnarray}\label{phmodevho}
|\Phi(t) \rangle \propto \sum_n c_n \exp \left( -i n
\sqrt{2JNgU_1} t \right) |n \rangle,
\end{eqnarray}
where $|n\rangle$ are the harmonic oscillator eigenstates and
where the coefficients $c_n$ depend on the exact dynamics of the
raising process. In particular, for symmetric initial conditions,
the phase distribution is always symmetric and only the even
eigenstates are populated. Then, the phase distribution breathes
with a frequency $2\sqrt{2JNgU_1}$, remaining always centered at
$\phi=0$. Moreover we notice that in such a case, the width in
the number distribution is not expected to be constant. 
In the variational ansatz treatment, we have to look again at
the orbits in the phase space.
The contour plot in Fig.~\ref{cplot}(b) is just to show how the
orbit modify from the limit of small $N$ to the limit of
large $N$. So let us consider  Fig.~\ref{cplot}(c).
The orbits around the minimum of ${\cal H}$ represent a time
evolution in which both the width of the number and phase
distribution change in time. The frequency of the small
oscillations around the equilibrium position can be calculted
analytically for $\exp(-\sigma_{\phi}^2/2)\sim1$ (see
Eq.(\ref{omegaplp}) for $U_2=0$) and coincides with the breathing
frequency of the phase distribution in the case of the
superposition of harmonic oscillator eigenstates if $gNU_1 \gg
J$. If this condition is not satisifed, one is not in the weak
coupling regime and the phase model is not valid.

The orbits in the $p$--$x_I$ space are characterised by very large oscillations 
in $p$.
Hence we cannot talk of frozen number fluctations. Nevertheless, with arguments 
similar to the ones used before, we can try to identify the orbit which describes
the dynamics at the end of the process. 
We estimate the maximum $p$ value in such an orbit to be
$p_M \sim p_{fin}$. The agreement with the numerical solution can be checked in
Fig.~\ref{posc} and it is within a factor of $2$.
The adiabaticity condition in this case consists in requiring that the final
state is superfluid, as the static solution would be. This means  that the 
minimum
value of $p$ corresponding to the same orbit as $p_{M}$  must be such that the
phase coherence is still good.
We found that a  final phase coherence corresponding to a minimum of 
$\lambda_+ = 1-\beta$ (with $\beta \ll 1/2$) is reached in 
process with typical time scales $\tau_{\beta} =2 \pi/8 JN\beta$.
Note that this condition is weaker than requiring  $p_{fin}=p_s$.
In fact  $\tau_{\beta} <  \tau_s =2\pi/\sqrt{8NJgU_1}$, since $\beta < 
\sqrt{gU_1/8JN}$,
as it would  correspond to the static solution.
This means that we still allow even big oscillation of $p$ around the 
equilibrium
value, as long as they do not destroy the phase coherence.
This corresponds to a breathing of the phase distribution which never becomes
completely smeared out. Moreover, while
$\tau_s$ depends only on the product $gN$, the adiabatic time scale 
$\tau_{\beta}$
scales like 1/N, getting easier and easier to be met for large $N$.

\subsubsection{Final superfluid phase}

\begin{figure}[htb]
\begin{center}
\epsfig{file=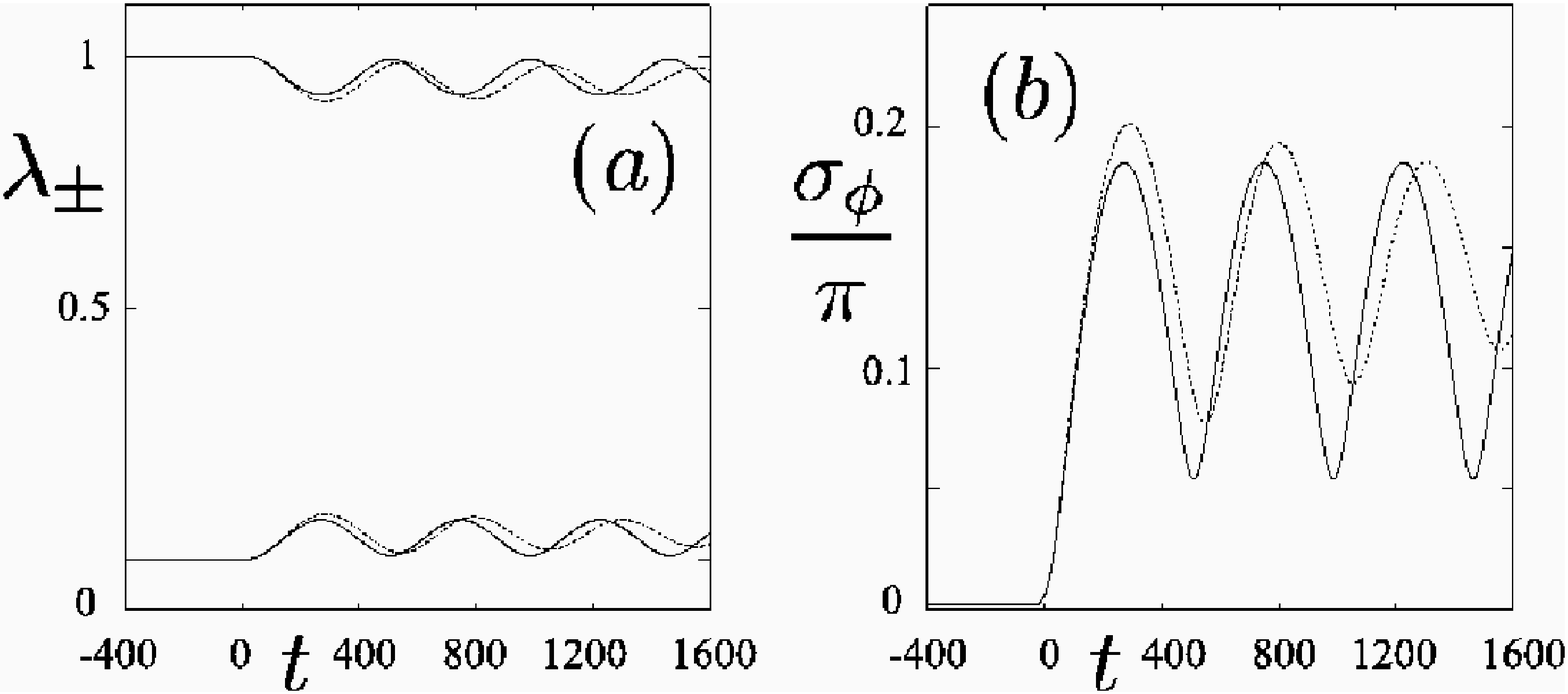,width=1\linewidth}
\end{center}
\caption{Eigenvalues $\lambda_{\pm}$ (a) and $\sigma_{\phi}/\pi$ (b). Numerical 
solutions
for the variational ansatz (full line) and for the phase model (dotted line);
$N=2 \times 10^5$, $g=0.005$ and $\tau=50$.}
\label{dyn3}
\end{figure}

As done before, we take now a particular case and check directly the
results with the one obtained in the phase model.
We choose $N=2 \times  10^5$, $g=0.005$ and $\tau=50$ and
find that  the eigenvalues $\lambda_{\pm}$ oscillate with frequency $\omega_p$
(see Fig.~\ref{dyn3}(a)):
$\lambda_+$ is always close to $1$ and $\lambda_-$ close to $0$.
This corresponds to the  breathing of the phase distribution
in the non negligible cosine potential or in the variational ansatz treatment
to one of the orbits in Fig.~\ref{cplot}(c).
No complete smearing out of the phase is observed (see Fig.~\ref{dyn3}(b)).
Those results are confirmed up to a good level by the phase model (see 
Fig.~\ref{dyn3}(a,b)),
even if  the oscillations of $\sigma_{\phi}$ are damped, due to the 
anharmonicities 
of the cosine potential.

The conditions that preserve the superfluid state are that the final tunneling
coupling $JN$ is comparable to the on-site interaction
and that the time scale of the process is  slow enough to allow  final small 
oscillations around the equilibrium.
To destroy the phase coherence in this parameter range, one should raise the 
barrier
faster in order to get larger oscillations, or raise it higher (case described 
in
Sec.~\ref{completesplit}).

\vspace{0.5cm}

To summarise, in this Section we have determined analytic expressions 
for the adiabaticity conditions for the internal dynamics,
i.e. we identified the time scale at which the barrier can be raised in order
to obtain the same relative phase properties as expected for the ground state.
Concerning a completed splitting process, a fragmented condensate without
any relative phase (no collapses and revivals) and characterised by
 a very narrow number distribution is reached in raising process with time scale
$\tau> 2\tau_r$.
This means that the process has to be slower for large $N$ and small $g$.
Instead in  an incomplete splitting process, the superfluid phase is preserved
for $\tau > 2 \pi/8 JN\beta$. This time scale becomes shorter for large $N$.
Both conclusions agree with the fact that in the Gross-Pitaevskii limit
($N\rightarrow \infty$ and $g\rightarrow 0$) the condensate is phase coherent.

\section{Discussion and conclusions}\label{concl}

We have solved  the two-mode model describing  the splitting of a condensate
by a potential barrier through a variational ansatz. We found coupling
between  the internal and  the  external dynamics of the mode functions.
We have identified the regimes in which the two dynamics decouple, and have
concluded that in the case of splitting starting from a condensate in
the ground state,
they do not influence each other in a dramatic way. Hence,
the internal and external excitations created by raising the barrier
can be estimated in a good approximation independently and have been 
characterised
as  a function of the interaction strength, the number of atoms and the time
scale of the process.

From our analytic estimations, confirmed by numerical resuts, we were able to 
identify the
time scales $ \tau_{z}$, $\tau_p $ and $\tau_{\beta}$, which define the 
adiabatic regime
for the external and  internal dynamics (respectively in the case of final
fragmented or phase coherent condensate)

\begin{eqnarray}
\tau_{z} \gg \frac{1}{\omega_z}, \;
\tau_p > 2 \tau_r =2  \frac{\pi}{ g U_1}, \;
\tau_{\beta} > \frac{ 2 \pi}{8 JN\beta}.
\end{eqnarray}
It is interesting to compare the adiabaticity condition for internal and 
external
degress of freedom in the case of final fragmented condensate, when the 
splitting
process can be consider to be completed. We normally found $\tau_{z}<\tau_p$;
 the case $N=200$, $g=5$ sets a boundary where internal and external degrees
of freedom enter simultaneously the adiabatic regime for $\tau \sim 20$
(see Fig.~\ref{posc}).
To get $\tau_{z}>\tau_p$ one needs
$gU_1$ to be large compared to the trapping frequency.
The quantity $gU_1$ is similar to the derivative of the
chemical potential with respect to the total number of atoms
 $\partial \mu / \partial N$. From a
Thomas-Fermi solution one gets that either in $1$ or $3$ dimensions, it scales
as a negative power of $N$ and a positive power of $g$. So one needs to have 
very
large $g$ or very small $N$, and our model fails in both limits.
Therefore, we claim that in the usual regime of many weakly interacting particles,
 the adiabaticity condition for the phase dynamics is more restrictive than the 
one
for the external dynamics.

We carried out a comparison of our model with the phase model, finding
a substancial very good agreement.
The numerical solution of the phase model consists in the integration of
a time dependent Schr\"{o}dinger equation. In practice, the number of 
wavefunctions
$\varphi_{m}(\phi)= \exp ( i m \phi) / \sqrt{2\pi} $ that one has to use
increases linearly with $N$, which leads to numerical problems.
In this sense for large $N$, the variational ansatz is more convenient and
has proved to give reliable results. Moreover the variational ansatz
treatment allows to include the external degrees of freedom in a natural way.

In the previous results we did not take asymmetries into account. It is
possible to include them in our ansatz, through the  unbalance in population
$m_0$ and through non symmetrically centered wave functions. In the case of
complete splitting, one ends up with a final constant unbalance in
population. The phase coherence shows the only new feature
that the center of the phase distribution is now drifting with a velocity
$\mu_1-\mu_2$, where $\mu_i$ are the chemical potentials of the two separate
condensates. A complete analysis of this case, which even consideres losses
and fluctuations in the total number of particles, can be found in
\cite{EPJD4-247}. Instead in the case of final phase coherent symmetric 
condensate,
the asymmetry can destroy the phase coherence. The final
unbalance in population might be  so big, that the wavepacket describing the 
phase 
distribution flies above the cosine potential: depending on the ``kinetic 
energy''
$gU_1 m_0^2$,
the cosine potential may become negligible and the same features (collapses,
revivals) as in the fragmented condensate case are observed. 

Possible  extensions of our model are the dynamic turning on of
an optical lattice,  where the initial harmonic trap
is deformed into a many wells potential. The instantaneous version of
such a process has recently been investigeted in \cite{PRA60-4902}.
Another problem of  great significance is the inverse process, i.e. the
merging of two condensates. This could allow to refill a condensate and
 be an important step towards a continuous atom laser.

\acknowledgments
This work was supported by the European Union TMR network ERBFMRX-CT96-0002 and
by the Austrian Science Foundation (Projekt Nr. Z30-TPH, Wittgenstein-Preis and 
SFB ``Control and measurement of Coherent Quantum Systems'').
C.~M. is grateful to Y.~Castin and A.~Sinatra for useful discussions, and  
thanks E.~Arimondo and G.~La~Rocca for comments.

\appendix
\section{Two--mode variational ansatz approach and phase model}\label{varanephmod}

Dealing with two coupled condensates, in this paper we have often
talked about  the number difference $m$ and
relative phase $\phi$.
In this appendix, we will define
number difference ${\hat m}$  and relative phase ${\hat \phi}$ 
operators, and derive the phase model Hamiltonian (\ref{phmodelham2}), 
discussing the approximations involved.

We first define the operators

\begin{mathletters}
\begin{eqnarray}
{\hat L}_x &=& \frac{1}{2} \left( a_1^{\dag}a_2 + a_2^{\dag}a_1 \right), \\ 
{\hat L}_y &=& \frac{i}{2} \left( a_1^{\dag}a_2 - a_2^{\dag}a_1 \right),  \\
{\hat L}_z &=& \frac{1}{2} \left( a_2^{\dag}a_2 - a_1^{\dag}a_1 \right),
\end{eqnarray}
\end{mathletters}
such that the usual angular momentum commutation relations are
fulfilled $[{\hat L}_i,{\hat L}_j]=i \varepsilon_{ijk}{\hat L}_k$.
After a small amount of algebra, the exact two--mode Hamiltonian in 
Eq.(\ref{expham}) can be rewritten as

\begin{eqnarray} \label{angmomham}
{\hat H} =g (U_1 -2 U_2)   {\hat L}_z^2
-2 J {\hat L_x} + 
 gU_2  \left( {\hat L}_x^2 - {\hat L}_y^2 \right).
\end{eqnarray}
In the subspace of fixed even total number of atoms $N$, the spectrum 
of ${\hat L}_z$ is given by all integer number in the interval $[-N/2,N/2]$.
The operator ${\hat L}_z$ coincides with the number difference 
operator ${\hat m}$. 
We have often  treated the eigenvalues  $m$ as  continuous,
but in general care should be exercised.
Given the phase operator ${\phi}$ such that $[{\hat \phi}, {\hat L_z}]=i$,
it is well--known that in the phase--representation 
${\hat L}_z=\hat m=-i\partial/\partial \phi$ and the eigenstates of ${\hat L}_z$
($\hat m$) with eigenvalue $m$ are 
$\langle \phi|m\rangle = \exp (im\phi) / \sqrt{2\pi}$ \cite{pascual}. 

The state of the system $|\Phi\rangle$ can be in
equivalent ways described as a superposition of eigenstates
of ${\hat m}$ ($c_m$ is the number distribution)

\begin{eqnarray}
 | \Phi \rangle = \sum_{m=-N/2}^{N/2} c_m |m\rangle.
\end{eqnarray}
or by a wave function in the $\phi$-representation given by

\begin{eqnarray}
\Phi(\phi)=\langle \phi | \Phi \rangle = 
\frac{1}{\sqrt{2\pi}}\sum_{m=-N/2}^{N/2} c_m \exp(im\phi).
\end{eqnarray}
Typical quantities characterizing $|\Phi\rangle$ are

- the width of the number distribution

\begin{eqnarray} \label{sigmam}
\langle {\hat m} \rangle &=& \sum_m m |c_m|^2, \nonumber \\
\sigma^2_m &=& \sum_m m^2 |c_m|^2 - \langle {\hat m} \rangle^2; 
\end{eqnarray}

- the width of the phase distribution

\begin{eqnarray} \label{sigmaphi}
\langle {\hat \phi} \rangle &=& \int_{-\pi}^{\pi} d \phi \, \phi  
\left| \Phi(\phi) \right|^2,  \nonumber \\
\sigma^2_{\phi} &=& \int_{-\pi}^{\pi} d \phi \, 
\phi^2 \left|  \Phi(\phi) \right|^2  - \langle {\hat \phi} \rangle^2
\end{eqnarray}
The uncertainty relations are  the same one as for angular momenta 
\cite{pascual}.

In order to write the Hamiltonian in the phase--representation, 
we  make some approximations which will lead to the simple phase 
model Hamiltonian and discuss under which conditions such approximations
are valid. 
We will describe the procedure only for the term ${\hat L}_x$, since  
for the term ${\hat L}_x^2-{\hat L}_y^2$ an analogous  one applies.
Using the raising and lowering operators ${\hat L}_{\pm}$, we write

\begin{eqnarray}
&& \langle \phi |2 {\hat L}_x |\Phi \rangle =  \sum_m c_m 
\langle \phi|\left( {\hat L}_+ + {\hat L}_- \right) |m\rangle =   \\
&&= \frac{N+1}{\sqrt{2\pi}}  \sum_m c_m 
\frac{1}{2}    e^{im\phi}  \nonumber    \\ 
&& \left[ \sqrt{1- \left( \frac{2m+1}{N+1} \right)^2 }  e^{i\phi} 
+ \sqrt{1- \left( \frac{2m-1}{N+1} \right)^2 } 
  e^{-i\phi}  \right].  \nonumber 
\end{eqnarray}
We assume a narrow and centered  number distribution (these assumptions 
will be quantified later on), so that we can expand the square roots
at first order in  $(m/N)^2$ and get

\begin{eqnarray}
&& \langle \phi |2 {\hat L}_x |\Phi \rangle =   \nonumber \\
&&=  (N+1) \cos \phi \Phi(\phi) + \nonumber   \\   
&& + 2 ( N+1 )   \left[ \frac{{\widetilde  {m^2}} +1/4}{(N+1)^2} \cos \phi  +
\frac{{\tilde m}}{(N+1)^2} \sin \phi \right] +    \nonumber   \\
&& + 
o\left( \frac{{\widetilde {m^4}}}{N^4}\right)  \approx  
N \cos \phi \; \Phi(\phi) , \label{cosphi}   
\end{eqnarray}
where we can roughly estimate
\begin{eqnarray}
&&\left|{\widetilde  {m^2}} \right|^2 \equiv \left|\sum_m c_m m^2 \langle \phi|m\rangle 
\right|^2 < \\
&&< \left|\sum_m |c_m|m^2 m^2 \right|^2 
 \leq \sum_m |c_m|^2 m^2 \sum_{m'} m'^2 = \nonumber   \\
&&= \left( \sigma_m^2 +\langle m \rangle^2 \right) 
\sum_{m=\langle m \rangle - \sigma_m}^{\langle m \rangle + \sigma_m} m^2 \sim  
\left( \sigma_m^2 +\langle m \rangle^2 \right)^2 \sigma_m < \nonumber \\
&&<\left( \sigma_m + |\langle m \rangle| \right)^4 \sigma_m,   \nonumber
\end{eqnarray}
so that $\left| \widetilde {m^2}\right| < (\sigma_m +| \langle m \rangle| )^2 \sqrt{\sigma_m}$
and where in a similar way
$|{\tilde m}| \equiv \left| \sum_m c_m m \langle \phi|m\rangle \right| \sim 
(\sigma_m +| \langle m \rangle| ) \sqrt{\sigma_m}$.
In an  analogous way,  the term ${\hat L}_x^2 - {\hat L}_y^2 $ gives

\begin{eqnarray} \label{cos2phi}
&&4 \left( {\hat L}_x^2 - {\hat L}_y^2 \right) \approx \\
&& \approx  N^2  \cos 2 \phi
-4\left[ (\widetilde {m^2}+1)  \cos 2 \phi + 2 {\tilde m} \sin 2 \phi \right] \approx \nonumber \\
&&\approx N^2 \cos 2 \phi.  \nonumber
\end{eqnarray}
All together the neglected terms have to be smaller than all the other 
terms in  the Hamiltonian, which implies

\begin{eqnarray}
&&  \left| \widetilde {m^2} \right| \sim  \left( \sigma_m  +|\langle m \rangle| \right)^2 
\sqrt{\sigma_m} \ll N^2 
\\
&&  \frac{2 J -gNU_2  }{N} \ll g (U_1-2U_2) . \label{valphmod}
\end{eqnarray}
For $|\langle m \rangle| \ll \sigma_m$, which is the typical situation
treated in this paper, from the first equation one gets
$\sigma_m \ll N^{4/5}$.
In the opposite limit where $|\langle m \rangle | \gg \sigma_m$
instead one has

\begin{eqnarray}
\frac{|\langle m \rangle|}{N} \ll \frac{1}{\sigma_m^{1/4}}.
\end{eqnarray}
For $\sigma_m \rightarrow 0$ this condition is not correct
and becomes simply $| \langle m \rangle | \ll N$.

To summarise, in the specific case treated in this paper, where no unbalance of 
population between the wells was assumed, the important condition of validity
for the phase model Hamiltonian is Eq.(\ref{valphmod}), which written for
$U_2=0$ takes the form

\begin{eqnarray}
J \ll gN U_1 /2.
\end{eqnarray}
We note that under such a condition, the ground state is characterised 
by $\sigma_{\phi} \agt  \sqrt{1/N}$ or equivalently $\sigma_m \alt \sqrt{N/4}$.

\section{Comparison between variational ansatz and phase model}\label{phmodevaran}

In this subsection we will compare the phase model with our variational
ansatz. For simplicity we set $U_2=0$, but the following discussion
could be repeated in the more general case. In particular for $U_2=0$
the phase model Hamiltonian reduces to the Josephson Hamiltonian
\cite{foundphys21-353}

\begin{eqnarray} \label{hphasemoldel}
{\hat H}=-g U_1 \frac{\partial^2}{\partial \phi^2} -JN \cos \phi.
\end{eqnarray}
The quantities $g U_1$ and $JN$ are respectively the on-site
energy splitting (charging energy) and the tunneling coupling
(Josephson coupling). It describe accurately the case of 
high barrier, where the two condensates are almost spatially
separated, leading to a negligible $U_2$, and are characterised
by very small number fluctuations, making the Josephson Hamiltonian
valid. The phase model Hamiltonian in
Eq.(\ref{hphasemoldel}) describes the motion of a particle in a
cosine potential and the classical limit is obtained for
$\sigma_m,\sigma_{\phi}\rightarrow 0$. In the following 
we discuss the classical and quantum limits comparing it
directly with the variational ansatz results.

In order to allow a complete comparison with the Hamiltonian 
in Eq.(\ref{hphasemoldel}), we 
now describe the coefficients $c_m$ in Eq.(\ref{sumc_m})
with four variational parameters $p$, $x_I$, $m_0$ and $\varphi$

\begin{eqnarray}
&&c_m= \\
&&{\cal N}(p)
\exp \left[- \left( \frac{1}{4 { p}}+i { x_I} \right)(m-m_0)^2 \right] \exp 
[im\varphi],  \nonumber
\label{coeffasy}
\end{eqnarray}
The variational parameters $x_I$ and $\varphi$ are the variables
conjugate to $p$ and $m_0$, allow the dynamic evolution and
vanish in the ground state. In the limit of broad number distribution
this ansatz describes a gaussian
superposition of number states centered in $m=m_0$. The
corresponding phase distribution $\Phi(\phi)$ is also a
gaussian, centered in $\phi=\varphi$. Notice that in the case of 
symmetric splitting treated before $m_0=0$ and $\varphi=0$ $\forall t$.

Replacing again the sum over $m$ from $-N/2$ to $N/2$ with an integral from 
$-\infty$ to $\infty$ (for $p \agt 1 $ and $x_I$ such that
$\sigma_{\phi} \alt \pi$), the widths of the number
and phase distribution are respectively given by Eqs.(\ref{sigmas})
and  the expectation values $ \langle a_i^{\dag} a_j \rangle$ and
$\langle a_i^{\dag 2} a_i^2 \rangle$ on the state $|\Phi
\rangle$ are now

\begin{mathletters} \label{expaaaasym}
\begin{eqnarray}
&& \langle \Phi | a_{1,2}^{\dag} a_{1,2} | \Phi \rangle =
\frac{N}{2} \mp m_0  , \\
&& \langle \Phi | a_{1,2}^{\dag2} a_{1,2}^2 | \Phi \rangle =
\frac{N^2}{4}   + p \mp N m_0 + m_0^2 , \\
&& \langle \Phi | a_1^{\dag} a_2 | \Phi \rangle =
\langle \Phi | a_2^{\dag} a_1 | \Phi \rangle^* =  \\
&& = \frac{N}{2} \exp \left(-\frac{\sigma_{\phi}^2}{2}\right)
\exp \left( i \varphi \right)  \nonumber \\
&& \hspace{2cm} 
\left[1-\frac{2p}{N^2} + \frac{2}{N^2} \left( m_0 -i2px_I \right)^2   \right].
\nonumber
\end{eqnarray}
\end{mathletters}

\subsubsection{Classical limit }

Let us first consider the classical limit and compare the
results obtained with the variational ansatz with the phase
model. We write the expectation value of the Hamiltonian
(Eq.(\ref{expham})) for $U_2=0$ in the classical limit by 
setting $\sigma_m=0$ and $\sigma_{\phi}=0$

\begin{eqnarray}
{\cal H}_{cl}= g U_1 m_0^2 - JN cos \varphi;
\end{eqnarray}
here we have assumed $m_0 \ll N$, neglected all terms $o(1/N)
\ll 1$ and all terms independent of $m_0$ and $\varphi$. One
sees immediately that this expression coincide with the
Hamiltonian in Eq.(\ref{hphasemoldel}), where the operators have
been substituted with their expectation values ($\langle {\hat
m} \rangle = m_0$ and $\langle {\hat \phi} \rangle = \varphi$).
So, the analogy with the phase model in the classical limit is
straighforward \cite{condmat9905059}.
In particular if $J> 0$ the stable equilibrium position is
$\phi=0$; for a kinetic energy such that $gU_1 m_0^2 < JN$ the
phase undergoes oscillations, otherwise if $gU_1 m_0^2 > JN$ the
phase flies above the cosine potential.

\subsubsection{Quantum limit}

In the quantum limit the width of the number and phase
distribution start to play an important role. The comparison
between the two models now is not so trivial, because we have on
the one hand a wavefunction in the phase representation,
$\Phi(\phi)$, and on the other hand $4$ variational parameters
($p$, $x_I$, $m_0$ and $\varphi$) which follow a classical
dynamics and should reproduce the features of $\Phi(\phi)$. We
consider here the simplified case of a wavefunction $\Phi(\phi)$
symmetric around $\phi=0$. In the variational ansatz picture,
this corresponds to $m_0=$ and $\varphi=0$ $\forall t$. This
last assumption is correct for symmetric initial conditions and
preserved at all times, how can be checked esplicitly in the
equations of motion.

Concerning the ground state of Hamiltonian (\ref{hphasemoldel}),
two different regimes can be identified: for $gU_1 \gg JN$, the
cosine potential can be neglected and the ground state is a flat
wavefunction, which corresponds to a completely undefined
relative phase between the two condensates. Instead for $gU_1
\ll JN$ the cosine potential is deep and the ground state is a
localised wavefunction, which describe a state with very well
defined relative phase.
In the variational ansatz approach, to find the ground state we
set $x_I=0$, $m_0=0$ and $\varphi=0$, and get
${\cal H}(p) = g U_1 p -J N \exp(-1/8p) \left[1-2p/N^2 \right]$.
For $gU_1 \ll JN$, the minimum of this Hamiltonian is $p=N/4$
corresponding to the Gross-Pitaevkii limit ($\sigma_{\phi} \sim
1/\sqrt{N}$) and if the hopping decreases $p\rightarrow 0$
($\sigma_{\phi} \gg \pi$), reproducing the same results as the
phase model. A quantitative comparison shows perfect agreement.

The time evolution was discussed already in Sec.~\ref{results},
where the analogy between the evolution of the phase
distribution $\Phi (\phi)$ and the evolution of the variational
parameters $p$ and $x_I$ was carried out.
The variational ansatz is able to predict all the main features
typical of the evolution of the phase distribution: collapses
and revivals of the relative phase for $gU_1 \gg JN$ and
breathing of the wavefunction in the cosine potential for 
$JN \gg gU_1$.
Moreover  the frequency of the small oscillations around the 
equilibrium point calculated in the variational approach 
concides with the splitting of the energy levels
of the phase model Hamiltonian  both in the limit of negligible
cosine potential and in the limit of deep cosine potential.

\vspace{0.5cm}

In this Appendix we have qualitatively compared the phase distribution
described by the phase model with the   
Gaussian ansatz for the coefficients of the state $|\phi\rangle$ in the
number representation. We have shown that in the classical limit
the expectation values $m_0$ and $\varphi$ of number and phase
(variational parameters indicating the centers of the respective
Gaussian distributions) are governed by the classical phase
model Hamiltonian. Moreover in the quantum description 
the parameters $p$ and $x_I$, related to the widths
$\sigma_m$ and $\sigma_{\phi}$, are able to reproduce the same
features of the time evolution predicted by the phase model, whose
limit of validity of the phase model (derived in App.~\ref{varanephmod})
is $J \ll gU_1N/2$.

We discussed the limit of fragmented condensate ($JN \ll gU_1$)
and the limit of single phase coherent condensate ($JN \gg
gU_1$). As already mentioned, for fixed $gN$, the limit
$N\rightarrow \infty$ corresponds to the Gross-Pitaevkii limit.
Infact, unless $J=0$, if $N\rightarrow \infty$ the condensate is
always phase coherent. For finite $N$, the above discussion
allows to judge whether the Gross-Pitaevskii description is
still valid or not.



\begin{references}
\bibitem{overlap} The definitions in Eqs.(\ref{u2},\ref{u3}) are not general: they 
are valid
always
is the static case when $\varphi_{1,2}$ are real and are always  true in the 
specific
case of our variational ansatz (see Eq.(\ref{variationalvarphi}))
\bibitem{b} We will make the following choice
$V_0=ab$ with $2b=1+2\log 2b \sim 1.75$, so that the potential wells keep as 
symmetric as possible
when $a$ varies.
This leads to oscillation frequencies of the same order of magnitude at the 
beginning
and at the end of the splitting process.
\bibitem{jav} This conclusion  is similar to the result in \cite{pra60-2351}. It 
is
difficult to make a direct comparison between the two, since due to the more
complex spatial and time dependence our tunneling rate does not decrease 
following
a simple exponential.
\bibitem{sigmaphiref} In  Fig.~\ref{dyn1},\ref{dyn2} (b) the values for 
$\sigma_{\phi}$ 
do not coincide for $\sigma_{\phi}/\pi \agt 0.58$, because we used 
Eq.(\ref{sigmasphi})
without correcting it, but they both indicate a complete smearing out of
the relative phase.
\end{references}
\end{document}